\documentstyle[prl,epsfig,aps]{revtex}   

\begin{document}

\draft


\title{A Model for Addition Spectra in Quantum Dots}

\author{Boris Shapiro\\
Department of Physics\\Technion-Israel Institute of Technology,\\
32000 Haifa, Israel}


\maketitle

\begin{abstract}

A simple model for addition spectra in quantum dots is proposed and studied.
 It is an extension
of  the standard charging model which assumes that the charge spreads
uniformely over the entire dot. The proposed model attempts to account
for a nonuniform distribution of the charge, by introducing an extra
  parameter
$U$. When $U$ increases, the distribution of the
conductance peak spacings changes from the Wigner-Dyson shape towards
a broader, more symmetric distribution.
\end{abstract}

\section{Introduction}
The combined effect of disorder and electron-electron interactions is an old
theme, to which Michael Pollak made important early contributions. The
topic is still very much alive. In particular, with the advance in
fabrication of mesoscopic electronic systems, the new feature of the
confinement, in one or more directions, has been added to the problem.

This paper deals with disordered quantum dots, i.e., small islands
of interacting electrons, in the presense of the random potential
of impurities. Let us denote by $E_0(N)$ the ground state energy of
a dot with $N$ electrons and define the addition spectrum as
$E_0(N+1)- E_0(N) \equiv \mu(N+1)$. This quantitiy is a subject of extensive
theoretical and experimental study. It is determined experimentally
by measuring the conductance of the dot, as a function of the gate
voltage $V_G$. The latter determines the electrostatic potential
on the dot.
 For a dot weakly coupled to the leads, and at low temperatures,
the conductance is negligible (Coulomb blockade), unless $V_G$
 is egual to $\mu(N)$, at which point the blockade is lifted and
an additional electron can enter the dot. Thus, the conductance
exibits a sequence of peaks. The distance between two consecutive peaks is
\begin{eqnarray}
\Delta_2(N)=\mu(N+1) - \mu(N) = E_0(N+1) -2E_0(N)+E_0(N-1).  
\end{eqnarray}

The simplest possible  description of the Coulomb blockade
is given in terms of  the charging model, also known as
 the constant-interaction model (for a recent review see Alhassid, 2000).
It is assumed in this model that the added, $(N+1)$-th, electron
occupies the $(N+1)$-th single particle state in the dot, with energy
$\epsilon_{N+1}$. The interaction between the added electron and each
of the $N$ electrons in the dot is described by a constant
potential $V_C$. These assumptions result in the following
expressions for the addition spectrum $\mu(N)$ and peak
spacing $\Delta_2(N)$:
\begin{eqnarray}
   \mu(N+1)= V_CN + \epsilon_{N+1},  \cr
   \Delta_2(N)=V_C +(\epsilon_{N+1}-\epsilon_N).
\end{eqnarray}
It follows from Eq. (2), that fluctuations in $\Delta_2(N)$ are
determined by the fluctuations in the single-particle level
spacing which, for a weakly disordered dot, are given by the
Wigner-Dyson distribution. Furthermore, if $N$ is odd, then the
$(N+1)$-th particle occupies a level which is alredy occupied
by a particle with opposite spin, so that $\Delta_2=V_C$. Thus,
according to the charging model, the distribution $P(\Delta_2)$
should exibit a bimodal structure, composed of a $\delta$-function
at $V_C$ plus a Wigner-Dyson function (shifted by $V_C$).
Experiments (Sivan et al., 1996, Patel et al., 1998, Simmel et al., 1999) 
, however, show an approximately Gaussian distribution
for $\Delta_2$, with a width considerably larger than the average
single-particle level spacing $\Delta$ and with no sign of a
bimodal structure. The disagreement between experiment and the
charging model led to a number of theoretical studies of $P(\Delta_2)$,
based on exact diagonalization of the microscopic Hamiltonian
 (Prus et al., 1996, Berkovits, 1998)
as well as on various approximations such as
 RPA (Blanter et al., 1997, Berkovits and Altshuler, 1997),
 the extreme classical limit (Koulakov et al., 1997)
and the self-consistent Hartree-Fock
 scheme (Walker et al., 1999, Cohen et al., 1999,
 Levit and Orgad, 1999), or the somewhat related
configuration interaction method (Benenti et al., 2000).
     In this paper we propose, and study, a simple model for the addition spectrum
 and the peak spacing  of a
weakly disordered (or chaotic) quantum    dot. The
electron spin is ignored, except for a few comments at the end of the paper.

\section{The Model}

$N$ interacting electrons in a quantum dot are described by the Hamiltonian

\begin{eqnarray}
H=\sum^N_{i=1} \left[ - \frac{\hbar^2}{2m} \bigtriangledown^2_i + V(\vec{r}_i)\right]
+ \frac{1}{2}\sum_{i<j} U(\vec{r}_i -\vec{r}_j) \ , 
\end{eqnarray}
where $m$ is the electron effective mass, $V(\vec{r}_i)$ includes the
confining potential as well as
 the random potential of
impurities, and $U(\vec{r}_i -\vec{r}_j)$ is the interaction potential. Let us discuss
first the two limits, of weak and strong interactions. In the first limit $\Delta_2$ is equal
to the single-particle level spacing  and, thus, obey the Wigner-Dyson
statistics. Small
corrections, due to interactions, can be treated
 in perturbation theory (Prus et al., 1996).

In the opposite case of very strong interactions (the "extreme classical limit")
 the potential energy
dominates , and the charges
 settle down in a certain configuration which minimizes the potential
energy of the system. For some special values of $N$,
 called ``magic numbers'', one can
expect particularly stable, i.e. low energy, 
configurations (Koulakov and Shklovskii, 1998, Morris et al., 1996).
 Specific values of the magic numbers
depend on the shape of the dot, the boundary conditions and the
type of interactions. For instance, for
a dot of an equilateral triangle shape these numbers
are $N=\ell(\ell + 1)/2$,
where $\ell$ is an integer. Indeed, for these values of $N$ one can view the
magic configuration inside the triangle as being cut out of an infinite Wigner crystal,
which is known to crystallize into a triangular lattice.
 In what follows, we consider a square dot,
with periodic boundary
conditions, often used in numerical calculations.
The triangular Wigner crystal, characteristic for an isotropic interaction,
 is clearly incompatible with the
square shape of the dot, and finding out the magic configurations for this geometry
becomes a formidable problem. However, for anisotropic interactions, or in the
presence of an underlying periodic potential, the charges can crystallize into
a square lattice. In this case the magic configurations should occur at
$N=\ell^2$. In order to keep the picture as simple as possible, without seriously
damaging the forthcoming reasoning, we shall stick to this example.

Consider, thus, a magic configuration of $N=\ell^2$ electrons, in a square dot
of size $L$.  The electrons inside the dot form a square lattice,
with a lattice constant $b=n^{-1/2}$, where $n=NL^{-2}$ is the
 electron concentration. The
lattice is pinned by the weak disorder in the dot. The main property
 of a magic configuration is its
stability, i.e., it can accomodate few extra particles without a major
rearrangement. An extra particle can cause some polarization
effects, by pushing
apart the charges in its neighborhood, but, basically, it has to reside in the
center of a cell, formed by 4 lattice charges. Furthemore,
the extra particles can
somewhat lower their energy by making quantum hops between neighboring cells.
Thus, the added particles can be viewed as "defects" moving
 on the intersticials of
a magic configuration. For an infinite perfect crystal  this kind
 of defects have been discussed long ago by Andreev and Lifshitz
(Andreev and Lifshitz, 1969) in their theory of quantum
 crystals (see also Tsiper and Efros, 1997). They
proposed that defects (or vacancies) can move through the crystal
coherently and can be treated as a gas of excitations. For a disordered
mesoscopic quantum dot this picture of "defects" becomes more
arguable. It is clear, in particular, that when too many particles
are added to a given magic configuration it is not appropriate to
speak in terms of many defects moving on top of that particular configuration:
one should rather speak about vacancies (holes) in the next magic
configuration. For instance, for $N=102$ one can envisage two defects
in the magic configuration $N=100$. However, for $N=119$ it is
clearly more appropriate to envisage two vacancies in the next
magic configuration, $N=121$ (compare to shells in atomic or nuclear physics).
Moreover, somewhere in between, say, at $N=110$ a major rearrangement
is likely to occur, and at this point it is hardly possible to
speak in terms of noninteracting defects or vacancies.
Inspite of its limitations,
 we will use this picture of defects, embedded in the
corresponding magic configuration, in order to formulate
  the forthcoming phenomenological model.

The main point of the above discussion is that in both limits, of
small and large $U$, the motion of a particle  added to
an interacting system is described in a single particle language:
 a bare electron, in the first case, and a "defect" in a magic
configuration, in the second. Next, we interpolate between the two
limits and arrive to the following, single particle, model
for the addition spectra in quantum dots.
The model is defined on a lattice with $2N$
sites. This lattice originates from the strong interaction limit, as described in
the previous paragraph. In this limit one sublattice ($A$) is occupied by the $N$
electrons of a magic configuration,
 while the other one ($B$) represents sites available for  extra
particles (Fig. 1). When an extra particle is placed on site $i$ of
sublattice $B$, the energy of the system increases by
$W_i +U_i$, where $W_i$ is the disorder related site energy, and
\begin{eqnarray}
U_i = \sum_{j\neq i} U(\vec{r}_i -\vec{r}_j) \
\end{eqnarray}
describes the interaction of the added electron with the electrons
in the dot.  Consider, as an example, a model potential (Walker et al. 1999)
consisting of a short range part $V_0\delta_{i,i+\eta}$ ($\eta$ designates the nearest
neighbors of site $i$) plus a constant term $V_C$, which describes the
charging energy due to the long range part of the interaction. For this
example $U_i=NV_C+4V_0$. More generally, one can replace
$4V_0$ by a phenomenological parameter $U$, whose relation to the pair potential
remains unspecified. Thus, in this "extreme classical limit", the $(N+1)$-th
particle, added to the $N$-particle magic configuration, will reside
on the site with the lowest value of $W_i$, on sublattice $B$. The
corresponding increase in energy is $\mu(N+1)= V_CN + U +W_i$. The
$(N+2)$-th particle will go into the site with the next lowest
value of $W_i$, etc. Similarly, one can remove particles from
a magic configuration and compare energies $E_0(N)$, $E_0(N-1)$, etc.
For instance, when one adds the $N$-th particle to the $(N-1)$-th
particle system, one fills in a vacancy in sublattice $A$. The main
point is that the energy $U_i$, corresponding to a site on
sublattice $A$, differs from that on sublattice $B$. This is because
a site on sublattice $A$ $(B)$ is surrounded by empty (occupied)
sites. We assign values $U$ and zero to sites on sublattices $B$
and $A$, respectively.

The full model for the fluctuations in addition spectra is obtained
by "switching on" the kinetic energy and, thus, interpolating between
the limits of small and large $U$. It is given by the matrix
\begin{eqnarray}
H_{ij} = W_i \delta_{ij} - t_{ij} + U_i \delta_{ij}  
\end{eqnarray}
where $t_{ij}$ accounts for hopping
between sites $i$ and $j$.   Energies $U_i$ assume two values:
$U$ for $i \in B$ and $0$ for $i \in A$. The model concentrates
on the fluctuations, i.e.,
 the charging energy
$V_C$, due to the long range part of the interaction, is not
included (it can be added to $\Delta_2$ at the end).
The matrix $H_{ij}$ is of size $2N \times 2N$. Its $(N+n)$-th eigenvalue
describes the energy (more precisely, its fluctuating part) of
the $n$-th particle added to the system of $N$ particles, so that
 $\Delta_2$ (with $V_C$ subtracted) is given by the difference
between a pair of consecutive eigenvalues.
  Let us stress that, although
the picture in the limit of very strong interactions, discussed above, was essential
for formulating the  model, this limit is quite remote from
the experimentally relevant situation: the latter corresponds to the
"liquid phase", rather than the solid one. In the liquid phase the
"magic numbers" do not play any essential role and the predictions of the model
are expected to be quite robust, i.e., independent of the specific
assumptions (e.g. a square lattice, rather than a triangular one)
 used in the "extreme classical limit". (If the number of electrons
is varied in too wide a range, one should allow for a weak dependence
of the parameters of the model on the number of electrons.)

 In the weak interaction limit, when
$U\rightarrow 0$, the model reduces to the Anderson model for a particle propagating
in a random potential. In this limit the lattice has no physical meaning and,
simply, provides discretization of the continuous space. Eq.~(5) defines a simple,
single-particle model for addition spectra. The model captures the right
physics in the two limits, of small and large $U$.  The weakness of the model is
that it is no more than a simple interpolation between the two limits.
 Such
an interpolation is, clearly, a big leap --- especially, since some
new quantum phases
might exist between the two limits
 (Chakravarty et al., 1999, Benenti et al., 1999). The model,
however, is not intended to describe subtle correlations in
 the ground state of $N$ interacting particles. Its only purpose is
to describe the energetics of added particles.
  In fact,
a rather similar picture --- with the same characteristic feature
of  two energy
bands in the large-$U$ limit, merging into one band for
 smaller $U$ --- is often used in
qualitative discussions of the Hubbard model (Mott, 1990). Below we
study the addition spectra
within the model of Eq.~(5).

\section{Numerical Results}

We assume that site energies, $W_i$, are uniformly distributed  in the interval
$[-W/2, \ W/2]$ and take $t_{ij}=t$ for nearest neighbors (and zero otherwise). $t$
is related to the particle mass by $t=\hbar^2/2ma^2$ where $a=(2n)^{-1/2}$ is
the lattice constant. This relation insures the correct value, $\Delta=2\pi\hbar^2/mL^2$,
for the single particle level spacing. The disorder strength is measured by the parameter $W/t$ and the interaction strength by the value of $U/t$.

Let us discuss in more detail the limit of strong interaction
and weak disorder, i.e., $W<t<<U$. The $U$-term in Eq.~(5) produces two levels, separated by
energy $U$.  Each of the levels is $N$-fold degenerate. The disorder and the
kinetic energy remove this degeneracy and broaden the levels into bands. The
lower band is occupied by the $N$ electrons, so that
 extra electrons must go into
the upper band. They reside in the low energy tail of that band, so that
the conductance peak spacings, $\Delta_2$, are given by the level spacings
in the tail.
   Since the density of states in the tail
 is much smaller than that in the middle of the band, one can expect
large (in units of $\Delta$) fluctuations in $\Delta_2$. Moreover, since the
tail states do not obey the Wigner-Dyson statistics, the same must be true for
the statistics of $\Delta_2$. Similarly, removing electrons from the
$N$-particle system amounts to creating holes in the high energy
tail of the lower band.
 Note that the width of the two bands is determined not by the hopping amplitude $t$ in the absence of interactions but, rather,
by an effective
 hopping amplitude $\tilde{t}\simeq t^2/U$. This is because hopping between
sites on sublattice $B$, such as sites 1 and 2 in Fig.~1, requires
visiting an intermediate
site (e.g. site 3), which belongs to the (occupied) sublattice $A$. Such
virtual visits are costly and contribute only in second order of the perturbation theory,
producing the effective hopping amplitude $\tilde{t}$ for particles (holes)
in the upper (lower) band. The estimate $\tilde{t}$ for the band width holds
only as long as $\tilde{t} > W$, i.e. $t > \sqrt{UW}$. In the opposite case
when $W < t < \sqrt{UW}$, the $t$-term in Eq.~(5) can be neglected. In this case the bands
have width $W$ and all eigenstates are localized.

Thus, the model in Eq.~(5) brings out in a simple way the origin of large fluctuations in the
peak spacings $\Delta_2$. The point is that, for large $U$, the model exhibits
two bands and the statistics of $\Delta_2$ is controlled by the states deep
in the tails of those bands. For small $U$ the two bands merge
into one band,
 and the statistics of $\Delta_2$ is determined by the single-particle levels
in the middle of that band. This qualitative picture is supported
 by numerics. As an
example consider a dot with a number of electrons near $N$=64.
 The situation is described, within our model, by a
lattice of 128 sites, i.e. by a matrix of size $128 \times 128$.
 Seperation between eigenvalues number $(N+i)$ and $(N+i-1)$ gives
the spacing between the corresponding conductance peaks (the eigenvalues
are ordered in energy). Let us look, for example, at the seperation
$\Delta_2$ between eigenvalues number 67 and 66.
  We take $t=1$ so that disorder and interactions are
described by the numbers $W$ and $U$ respectively.
 The distribution of $\Delta_2$,
for an ensemble of 3000 matrices, for $W=4$ and various values
 of $U$ is shown in Fig.~2 ($\Delta_2$ is
normalized to $\Delta$, so that the $x$-axes is labeled by
$\delta$=$\Delta_2$/$\Delta$). For $U=0$ the distribution is close to
Wigner-Dyson. This fact demonstrates that $W=4$, although not small compared to
unity, corresponds to the diffusive regime (to reach diffusion behavior with
small $W$ one would need larger matrices, i.e., larger number of electrons in the dot).  For $U$ near 2 significant deviations from
Wigner-Dyson statistics start to show up, and between $U\approx 2$ and $U\approx 3$ the
distribution rapidly changes from Wigner-Dyson towards a more
 symmetric, broader distribution. For
still larger values of $U$ (roughly, larger than 10) the distribution
  gradually
approaches a limiting Poissonian shape. The origin of this limiting
 distribution is clear:
as was explained above, for sufficiently
 large $U$ the $t$-term in Eq.~(5) can be
neglected, and one ends up with an ensemble of diagonal
 matrices $H_{ij}=(W_i + U_i)\delta_{ij}$, with statistically independent
eigenvalues.
Histograms similar to those in Fig. 2 can be obtained also for other
pairs of consecutive eigenvalues, not too far from the 64-th
eigenvalue, $\epsilon_{64}$ (otherwise one should change the size of
the matrix, adjusting it to the nearest magic number). This eigenvalue
is somewhat exceptional, in the sence that it is
between $\epsilon_{64}$ and $\epsilon_{65}$
that a gap opens up, for a sufficiently large
$U$ (see Fig. 3, for $U$=5). Therefore, for large $U$, the distribution of $\Delta_2$ for this
case is shifted by $U$ (Fig. 4). The other special feature of this
particular case is that, for very large values of $U$, the distribution
assumes a semi-Poisson shape (a semi-Poisson distribution is defined
as $p(x)=xe^{-x}$).

The average density of states, for the same ensemble of matrices,
 is plotted in Fig.~3. For $U=0$ one has the single band of the Anderson model. When $U$
increases, a gap starts to form in the middle of the band and, eventually, two separate
bands emerge. There is a clear correlation between formation of the gap and
deviation of the peak spacing statistics from the Wigner-Dyson shape. Note that
significant deviations from Wigner-Dyson statistics develop long before a genuine gap oppens up in the spectrum, i.e., standard metallic conductivity can
coexist with "non-metallic" thermodynamic properties.

\section{A Random Matrix Model}.

One can further simplify the model, replacing the Anderson matrix, i.e., the
first two terms in Eq.~(5), by a $2N \times 2N$ Gaussian
 orthogonal ensemble (GOE) (Mehta, 1991).
This leads to a matrix model
\begin{eqnarray}
H = G+A 
\end{eqnarray}
where $G$ is taken from the GOE and $A$ is a $2N \times 2N$ (non-random) diagonal matrix, with
half of the eigenvalues equal to $U$ and the other half equal to zero.
It is interesting that the same
  matrix model can be arrived
at by entirely different, more abstract arguments. Let's start with
a standard Gaussian matrix, $G$, which is known to give a good description
of the single particle level statistics in a chaotic or disordered
quantum dot. The matrix ensemble $G$ has a $O(M)$ symmetry, where $M$
is the number of states (orbitals) spanning the relevant Hilbert
space. Consider now $N$ interacting particles in the dot and employ
some self-consistent description, e.g. the Hartree-Fock scheme. The
outcome will be $M$ self-consistent orbitals, $N$ of which are occupied
and the rest, $M-N$, are empty. One can make orthogonal transformations
{\em seperately} among the $N$ occupied and the $M-N$ empty orbitals,
but not between the two types of orbitals. Thus, the symmetry is broken
from $O(M)$ to $O(N) \times O(M-N)$. The matrix $A$ in  Eq.(6) implements
such symmetry breaking and introduces the two types of orbitals.
Thus, the matrix model of Eq. (6) seems to emerge from some
very general arguments.

Let us study the statistics of $\Delta_2 = \epsilon_{N+i}-\epsilon_{N+i-1}$,
where $\epsilon_n$ is the $n$-th (ordered) eigenvalue of $H$ and
$i$ is a small integer.
Qualitatively, the picture is similar to that outlined above, i. e.,
 under the increase of $U$, the distribution $P(\Delta_2)$ crosses over from the Wigner-Dyson shape to a
broader, more symmetric  distribution (Fig. 5). The crossover is related to the
formation of a gap in the spectrum of $H$ (Fig. 6). Eventually, for
sufficiently large $U$, the Wigner semicirle of the GOE splits
into two well separated semicircles, and $P(\Delta_2)$ is controlled
 by the tail states, where the density of levels is extremely low
(by a factor $N^{1/3}$ smaller than in the middle of the band (Mehta, 1991)).
It is interesting to note that Brezin and Hikami (Brezin and Hikami, 1998) 
have studied the matrix model in Eq.~(6), but with $G$ being the Gaussian unitary ensemble, instead of the GOE. They considered the "critical situation", when the
two bands just touch one another, and showed that a new universal level
statistics emerge in the large-$N$ limit.

\section {Conclusions}

 A simple model for addition spectra is proposed and studied.
 It can be viewed as an extension of the charging model. The latter is
based on the assumption that the added particles spread over the entire
dot, charging it uniformly. This assumption is, clearly, inadequate in
the limit of  strong interactions, when the charge density
 in the dot exibits strong spatial changes.
 The proposed model tries to account for this feature
 in the most crude phenomenological way, by
introducing the parameter $U$. This turns out to be sufficient for
obtaining a qualitatively correct behavior for the conductance peak
spacings. The model, or rather its version in Eq.(6), is also of interest
in the random matrix theory. The point is that the standard
(orthogonal or unitary) Gaussian
ensembles describe well the addition spectra in a disordered dot
only in the absence of interactions. The interaction breaks the
$O$- or $U$-invariance of a Gaussian ensemble. This is seen most
clearly in the "extreme classical limit", when the electrons become
localized on particular site-orbitals. Matrix $A$ in Eq.(6) implements
 such symmetry breaking. There are, of course, other ways to break
this symmetry, some of which lead to a crossover between the Wigner-
Dyson and the Poisson statistics for level spacings 
(for a review see Shapiro, 1996). It would be,
perhaps, of some interest to  investigate the connection between
such generalized matrix ensembles and the addition spectra in disordered
quantum dots.

The model in Eq.~(5) was designed for spinless electrons. For spinfull electrons
the extra particles can hop on occupied sites, so that there is no need for the second
sublattice. The "minimal" model is defined on a $N$-site lattice,
 corresponding to the the $N$-electron magic configurations,
which is the same as for the spinless case. However, each site of
the lattice now contains 2 orbitals. They correspond to single
and double occupation, respectively, and differ by energy $U$.
For zero $U$ one has a single band, with doubly degenerate levels. The statistics of $\Delta_2$ exibits in this case a bimodal structure, characteristic for
non-interacting electrons with spin. For large $U$ the picture
 is essentially the same as for the spinless case:  extra particles (holes
) go into the
tail of the upper (lower) Hubbard band (Mott, 1990). Since there is
 no spin degeneracy in these bands (they correspond to different
 site-occupation nubmers), there is no trace of a bimodal structure.
The absence of such structure, as well as other spin effects, have been
studied earlier, both for disordered and clean 
quantum dots (Berkovits, 1998, Hirose and Wingreen, 1999, Baranger et al.,
2000, Cha and Yang, 2000, Jacquod and Stone, 2000).

\section*{Acknowledgements}

I benefited greatly from discussions with A. Auerbach, J. Feinberg,
 M. Janssen,
A. Kamenev, Y. Levinson, G. Montambaux, J.-L. Pichard, R. Pnini,
 M. Pollak and U. Sivan. These discussions were instrumental in
shaping the final version of the model. I am greatful to D. Gangard and,
in particular, to A. Retzker for the help with numerics and figures.
 The research was partially supported by the DIP project
on "Quantum Electronics in Low-Dimensional Systems" and by
the Fund for promotion of research at the Technion.
\pagebreak

\begin{center}
\section*{References}
\end{center}
\begin{itemize}

\item[] Alhassid, Y., 2000, Rev. Mod. Phys., {\bf 72}, 845.

 \item[] Andreev, A.F., and  Lifshitz, A.M., 1969, Zh. Eksp. Teor. Fiz. {\bf56},
 2057 [Sov. Phys. JETP {\bf29}, 1107]

 \item[]Baranger, H.U., Ullmo, D., and  Glasman, L.I., 2000,
 Phys. Rev. {\bf B61}, R2425.

\item[] Benenti, G., Waintal, X., Pichard, J.-L., and  Shepelyansky, D., 2000,
  Europ. Phys. Jour. {\bf B17}, 515.

 \item[] Benenti, G., Waintal, X., and Pichard, J.-L., 1999,
Phys. Rev. Lett. {\bf 83}, 1826.

\item[]  Berkovits R., 1998, Phys.\ Rev.\ Lett. {\bf 81}, 2128.

\item[]  Berkovits, R., and  Altshuler, B.L., 1997, Phys.\ Rev.\ {\bf B55}, 5297.

\item[] Blanter, Ya.M., Mirlin, A.D., and Muzykantskii, B.A., 
1997, Phys.\ Rev.\ Lett.
{\bf 78}, 2449.

 \item[] Brezin, E., and  Hikami, S., 1998, Phys.\ Rev.\ {\bf E58}, 7176.

\item[]Cha, M.C., and Yang, S.R.E., 2000, Phys. Rev. {\bf B61}, 1720.

 \item[] Chakravarty, S., Kivelson, S., Nayak, C. and  Voelker, K., 1999, Phil.Mag.
 {\bf B79}, 859.

\item[] Cohen, A.,  Richter, K. and  Berkovits, R., 1999, 
Phys.\ Rev.\ {\bf B60}, 2536.

 \item[]Hirose, K. and Wingreen, N.S., 1999, Phys. Rev. {\bf B59}, 4604.

\item[] Jacquod, P.,
and  Stone, A.D., 2000, Phys. Rev. Lett. {\bf 84}, 3938.

\item[] Koulakov, A.A., Pikus, F.G. and  Shklovskii, B.I., 1997, Phys.\ Rev.\ {\bf

B55}, 9223.

\item[] Koulakov, A.A. and Shklovskii, B.I., 1998, Phys.\ Rev.\ {\bf B57}, 2352.

\item[] Levit, S. and  Orgad, D., 1999,  Phys.\ Rev.\ {\bf B60}, 5549.

\item[]  Mehta, M.L., 1991, Random Matrices (Academic Press).

 \item[]  Morris, J.R., Deaven, D.M., and Ho, K.M., 1996,
 Phys.\ Rev.\ {\bf B53}, R1740.
 
\item[] Mott, N.F., 1990, Metal-Insulator Transitions (Taylor$\&$ Francis).

\item[] Patel, S.R., Cronenwett, S.M., Stewart, D.R., Huibers, A.G.,
Markus, C.M., Doruoz, C.I., Harris, J.S., Campman, K., and Gossard, A.C.,
1998,  Phys.\ Rev.\ Lett. {\bf 80}, 4522.

\item[]  Prus, O., Auerbach, A., Aloni, Y., Sivan, U., and
Berkovits, R., 1996, Phys.\ Rev.\ {\bf B54}, R14289.

 \item[] Shapiro, B., 1996, Int. Jour. Mod. Phys. {\bf B10}, 3539.
 
\item[] Simmel, F., Abusch-Magder, D., Wharam, D.A.,
Kastner, M.A., and Kotthaus, J.R., 1999, Phys.\ Rev.\ {\bf B59}, R10441.

\item[] Sivan, U., Berkovits, R., Aloni, Y.,
Prus, O., Auerbach, A., and Ben-Yoseph, G., 1996,
 Phys.\ Rev.\ Lett. {\bf 77}, 1123.

 \item[]  Tsiper, E.U., and  Efros, E.L., 1997, J. Phys. Cond. Matter {\bf 9},
 L561.
 
\item[] Walker, P.N., Montambaux, G., and Gefen, Y., 1999,
 Phys.\ Rev.\ {\bf B60}, 2541.

\end{itemize}

\pagebreak

\begin{figure}

\caption {The large-$U$ limit of the model.
 In a magic configuration of $N$ electrons sublattice $A$
is occupied (full circles).
  Extra electrons can propagate on sublattice $B$ (empty circles).
Vacancies can be created by removing particles from sublattice $A$.
Hopping between sites 1 and 2 on sublattice $B$ occurs
via occupied sites, like site 3.}

\label {Fig.1}

\end{figure}

\begin{figure}

\caption {The distribution of conductance peak spacings,
 $P(\delta)$, is plotted for various
values of $U$, between 0 and 5. The seperation, $\Delta_2$,
between eigenvalues number 67 and 66 is measured
in units of $\Delta$, i.e., $\delta$=$\Delta_2$/$\Delta$. The constant
charging energy, $V_C$, is not shown in this plot. It can be accounted for
by shifting all the curves by $V_C$.
 The dotted curve is the Wigner-Dyson distribution.}

\label {Fig.2}

\end{figure}

\begin{figure}

\caption {The average density of states, $\nu$, for the same
values of $U$
 as in Fig. 2. Energy (the horisontal axes) is measured in units of  $t$.}

\label {Fig.3}

\end{figure}

\begin{figure}

\caption {The same as Fig. 2 but for the seperation between eigenvalues
65 and 64. It is between these eigenvalues that a gap opens up, in
the large-$U$ limit.}

\label {Fig.4}

\end{figure}

\begin{figure}

\caption {Distribution for  the seperation between eigenvalues
67 and 66, for the matrix ensemble defined in Eq. (6),
for values of $U$ between 0 and 2.5. The seperation
is measured in units of the average level spacing at $U=0$, i.e.,
in the absence of matrix $A$. The matrix size is $128 \times 128$.
 The size of the ensemble is 3000.}

\label {Fig.5}

\end{figure}

\begin{figure}

\caption {The average density of states, $\nu$,
of the matrix ensemble, for the same values of $U$ as in Fig. 5.}

\label {Fig.6}

\end{figure}
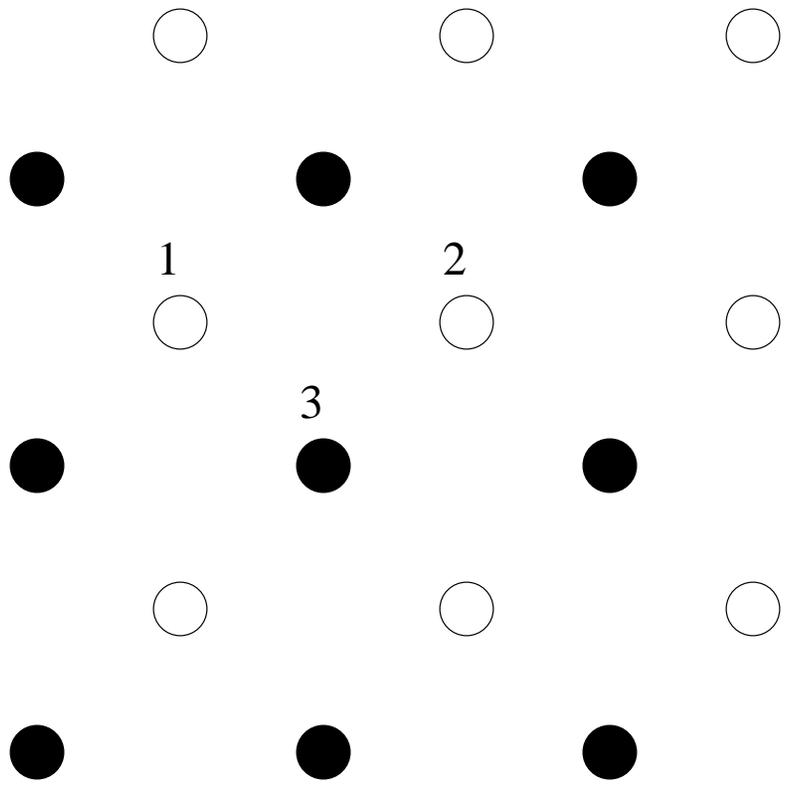
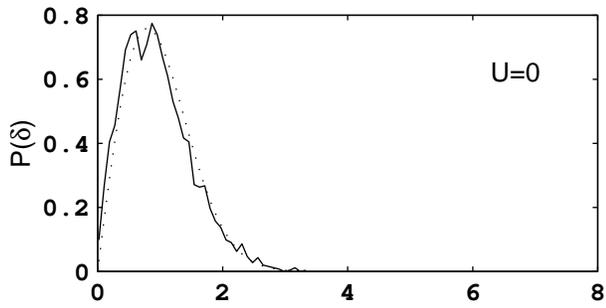
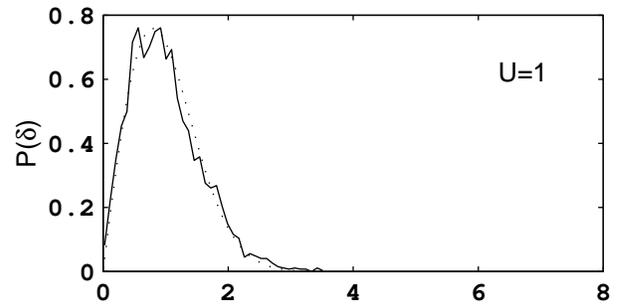
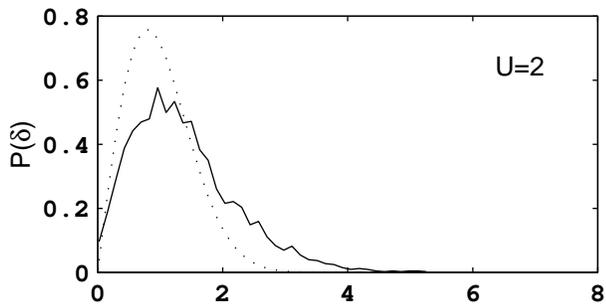
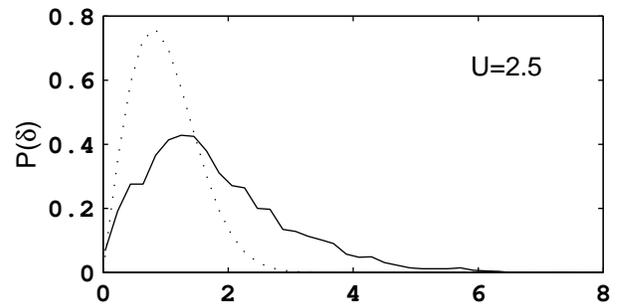
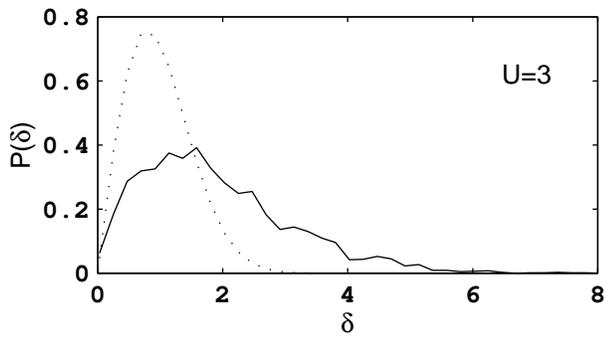
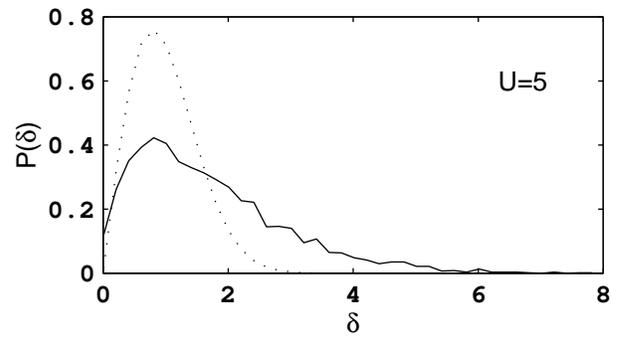
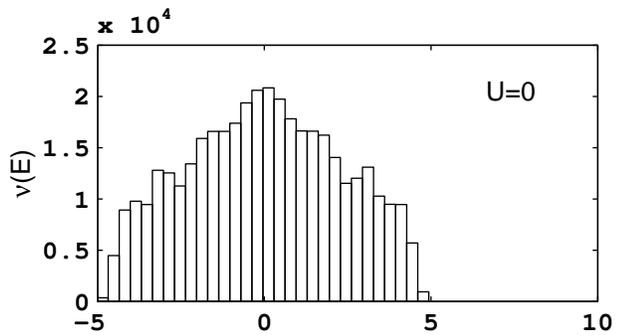
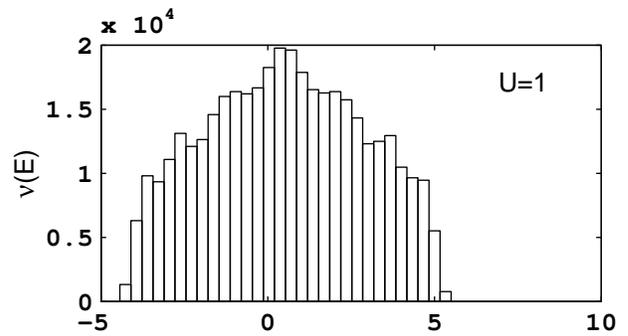
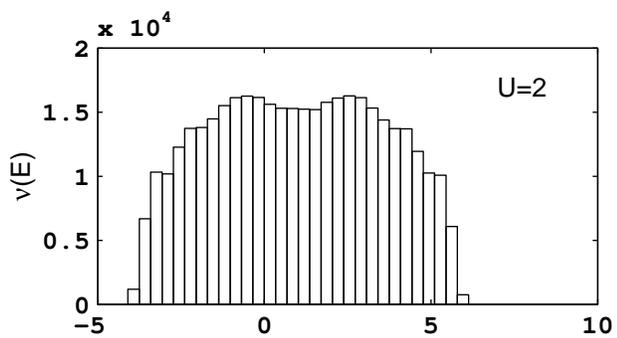
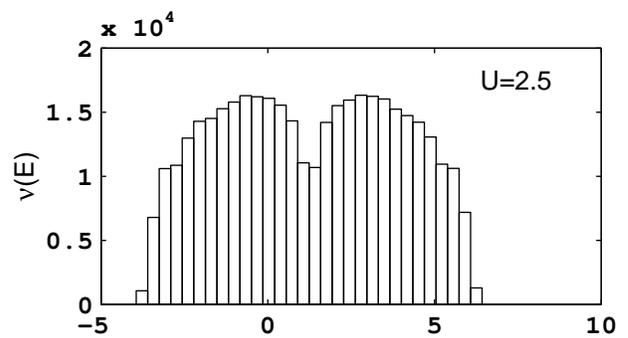
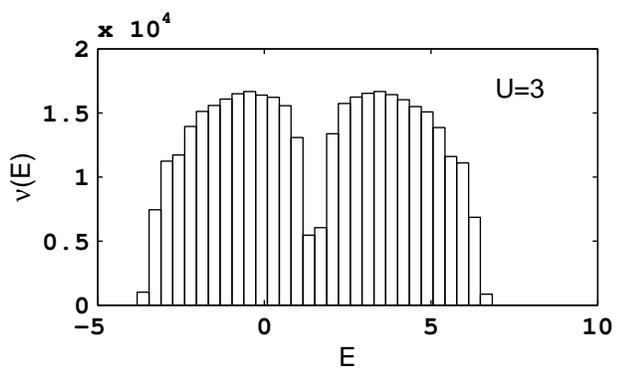
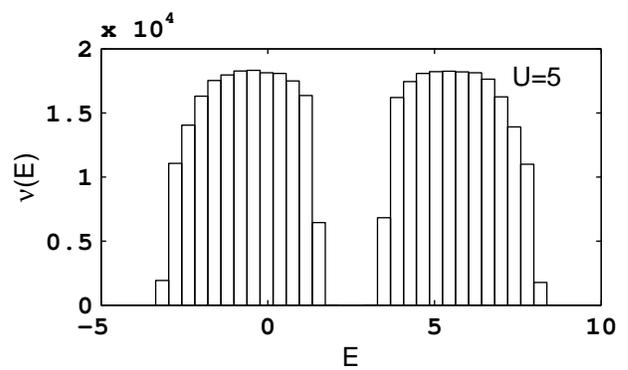
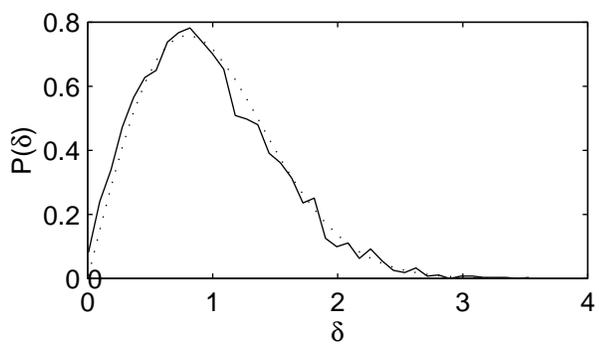
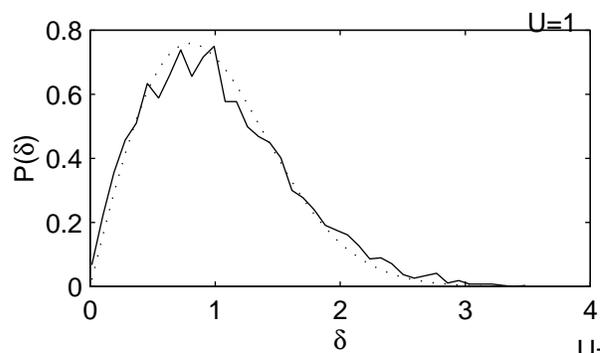
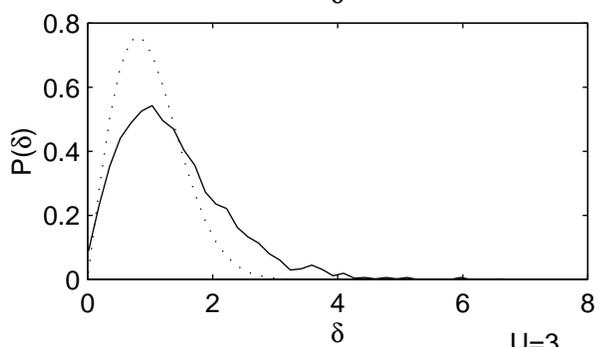
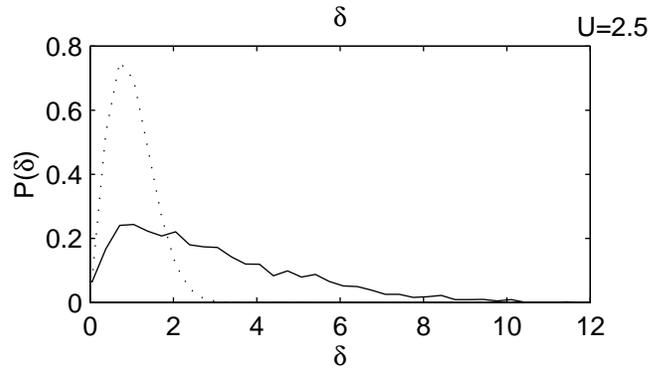
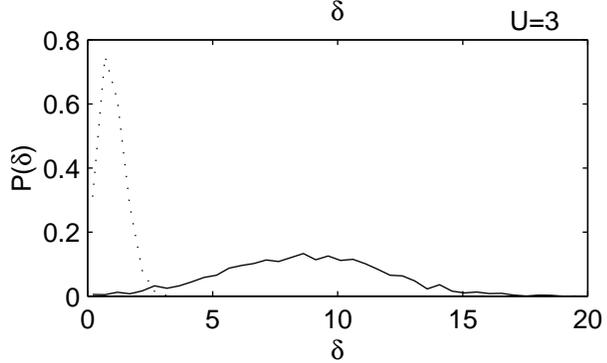
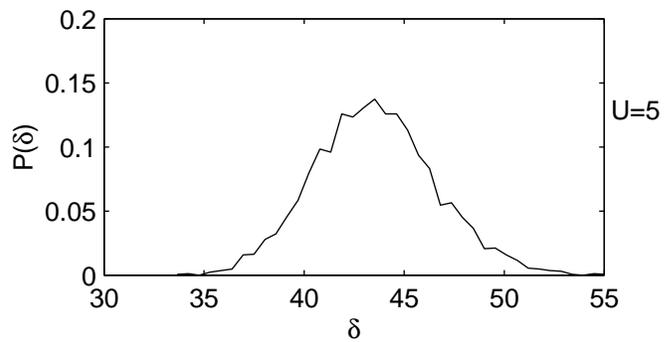
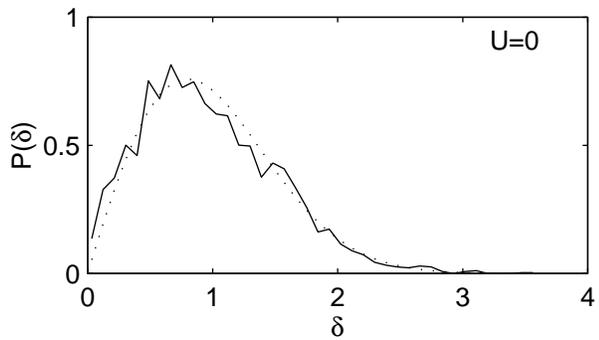
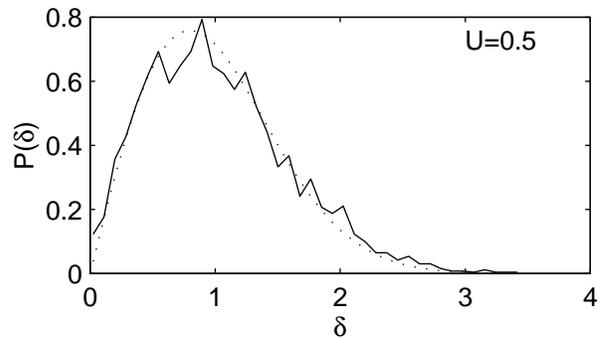
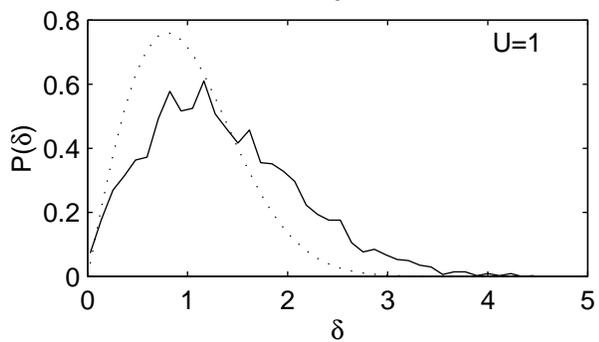
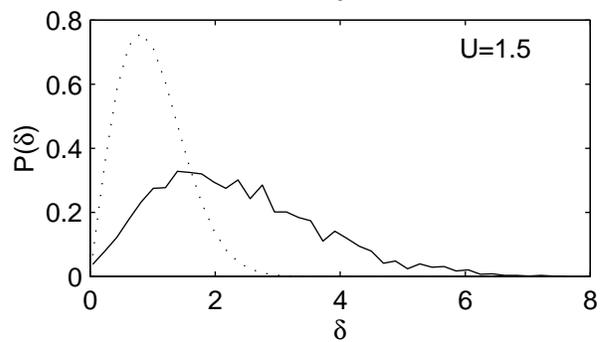
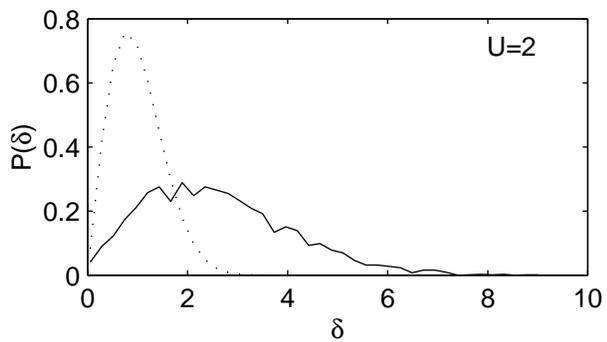
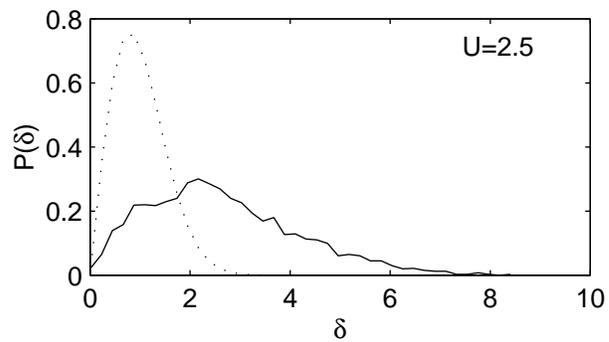
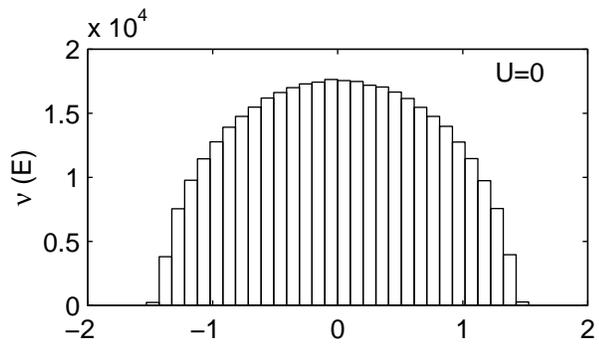
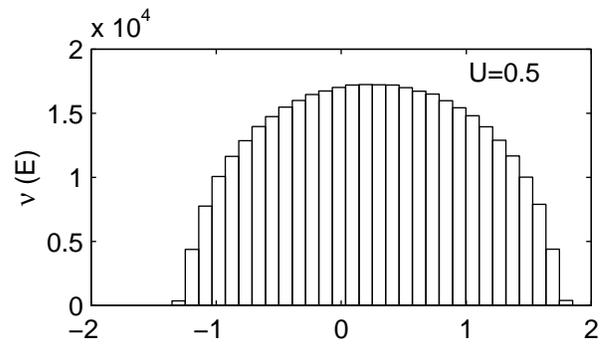
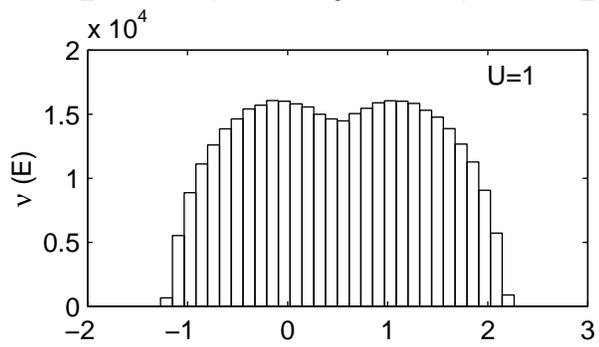
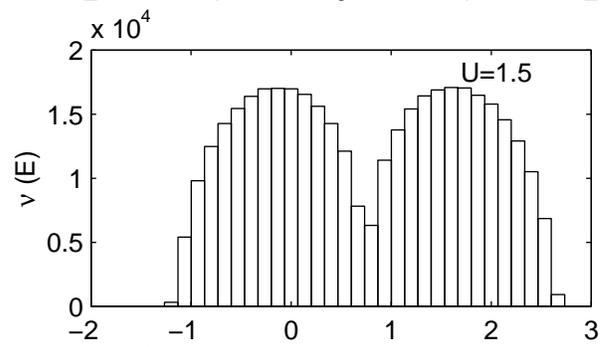
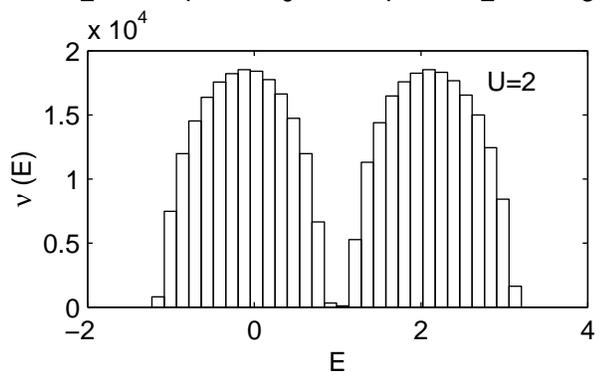
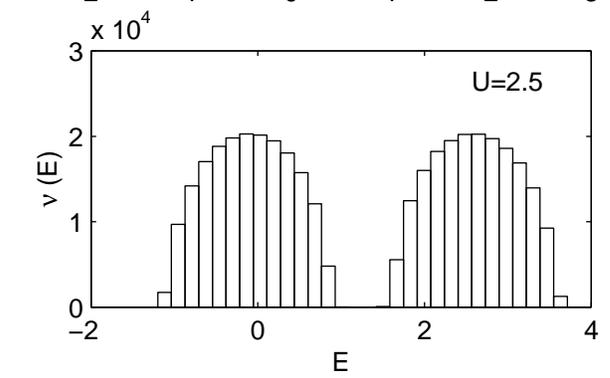
\end{document}